\documentclass[aps,prl,twocolumn,longbibliography,amsmath,amssymb,superscriptaddress]{revtex4-1}
\usepackage{amsmath} 
\usepackage {units}
\usepackage{here}
\usepackage{graphicx} 
\usepackage{verbatim} 
\usepackage{color} 
\usepackage{bm}
\usepackage{subfigure} 
\usepackage{hyperref} 
\usepackage[normalem]{ulem} 
\usepackage{siunitx} 

\begin{document}
\title{Negative autoregulation controls size scaling in confined gene expression reactions}
\author{Yusuke T. Maeda}\email{ymaeda@phys.kyushu-u.ac.jp}
\affiliation{Department of Physics, Kyushu University, Motooka 744, Fukuoka 819-0395, Japan}
\date{\today}

\begin{abstract}
Gene expression via transcription-translation is the most fundamental reaction to sustain biological systems, and complex reactions such as this one occur in a small compartment of living cells. There is increasing evidence that t physical effects, such as molecular crowding or excluded volume effects of transcriptional-translational machinery, affect the yield of reaction products. On the other hand, transcriptional feedback that controls gene expression during mRNA synthesis is also a vital mechanism that regulates protein synthesis in cells. However, the excluded volume effect of spatial constraints on feedback regulation is not well understood. Here, we study the confinement effect on transcriptional autoregulatory feedbacks of gene expression reactions using a theoretical model. The excluded volume effects between molecules and the membrane interface suppress the gene expression in a small cell-sized compartment. We find that negative feedback regulation at the transcription step mitigates this size-induced gene repression and alters the scaling relation of gene expression level on compartment volume, approaching the regular scaling relation without the steric effect. This recovery of regular size-scaling of gene expression does not appear in positive feedback regulation, suggesting that negative autoregulatory feedback is crucial for maintaining reaction products constant regardless of compartment size in heterogeneous cell populations.
\end{abstract}
\maketitle

\section*{Introduction}

A micron-sized compartment that separates the cytoplasmic space from the exterior environment is a fundamental feature of living cells \cite{noireaux1,jewett}. DNA, which stores genetic information, is encapsulated in a tiny cellular compartment with a lipid membrane. Catalytic proteins are synthesized by the transcription of the genetic information stored in the DNA sequence into messenger RNA (mRNA) and the translation of the mRNA sequence into a single chain of amino acids. In bacteria, these complex transcription-translation (TXTL) reactions proceed autonomously under cell-sized confinement of a few microns \cite{noireaux2,noireaux3,macdonald,hibi,lu}. In contrast, protein complexes that regulate TXTL reactions, such as the ribosomes, have a finite size of a few tens of nanometers \cite{milo}. In particular, microorganisms such as bacteria densely enclose proteins, mRNA, and DNA within the cytoplasm in tiny cell bodies. In such micron-sized capsules, the ratio of the surface layer to the total volume is large, making the effect of finite molecular size negligible. The impact of such excluded volume effects has been reported in recent studies\cite{tan,norred,Gonzales,vibhut}. With regard to in vitro systems, the addition of crowding agents such as an inert polymer in cell-free extracts induces the crowding of large protein complexes involved in gene expression, and this results in an enhancement of gene expression. In other words, the finite size of molecules is closely related to the control of intracellular reactions. Moreover, physical effects occur not only between molecules but also between molecules and boundaries \cite{sakamoto,garenne2}. Hence, the question of how TXTL reactions confined to a small compartment are affected is vital for understanding the physical nature of confined gene expression \cite{barziv,ziane1} and for implementing cell-free biochemical reactors enclosed in a space as small as bacteria \cite{maeda1,noireaux4,noireaux7,rivas,kato}.

Previous studies have examined gene expression in cell-sized water-in-oil emulsions as artificial cells to understand the excluded volume effect in confined TXTL reactions. This approach has shown that TXTL reactions can be suppressed in small artificial cells and that the amount of protein expression is not proportional to the volume of the artificial cells \cite{sakamoto}. In contrast, the amount of protein expression in large artificial cells increased proportionally with the volume of the artificial cells. Such anomalous size dependence in confined TXTL reactions suggests that the excluded volume effect significantly suppresses protein production under spatial constraints. It should be noted that confinement-induced repression must be resolved to construct a biochemical factory utilizing cell-free TXTL reactions \cite{barziv,ziane1,rivas}. Many techniques have been developed to create artificial cells of uniform size using droplet generators and to encapsulate artificial cell reactors in devices. However, as these technologies advance and become smaller and more precise, it will be necessary to construct reaction systems that consider steric effects as well. A remaining challenge is to explore the design of scalable cell-free reactors that reduce suppression due to finite-size effects of molecules and achieve stable gene expression from the submicron-sized reactor to the test tube. It remains to be seen what mechanism is needed in confined TXTL reactions to sustain ordinary size-dependence.

The key to addressing this issue is the regulatory network in the TXTL reactions, in which the amount of expressed proteins is controlled by transcriptional factors \cite{savageau}. For instance, the autoregulatory feedback where the transcription factor regulates its encoding gene has been identified widely as a ``network motif" in gene regulatory networks \cite{alon1,alon2}. In particular, negative autoregulatory feedback (NAF) control is an abundant network motif. NAF control has broad functions, fast kinetic response \cite{alon3}, suppressing concentration variability \cite{serrano,simpson}, mutational robustness \cite{brem} and the protein synthesis on demand \cite{murray}. Although there are extensive studies on NAF control in gene regulatory networks in bulk, its regulatory role in confined TXTL reactions is not well understood. 

In the present study, we investigate confined TXTL reactions with NAF in a cell-sized compartment by considering a mathematical model. We analyze the size dependence of the amount of protein expressed with the NAF control. The mathematical model shows that the anomalous size-dependent scaling is dampened in the presence of NAF control at the transcriptional level. Such size scaling approaches are close to ordinary volume dependence because mRNA synthesis is suppressed by the excluded volume effect in the small compartment and by the action of NAF control in the large compartment. Our findings may provide insights into the functional role of NAF control in the homeostasis control of the TXTL reaction under the variability of cell size.

\section*{Results}

\subsection*{Gene expression reaction in a confined space}
\begin{figure}[tbp]
\centering
\includegraphics[scale=0.26,bb=360 0 680 452]{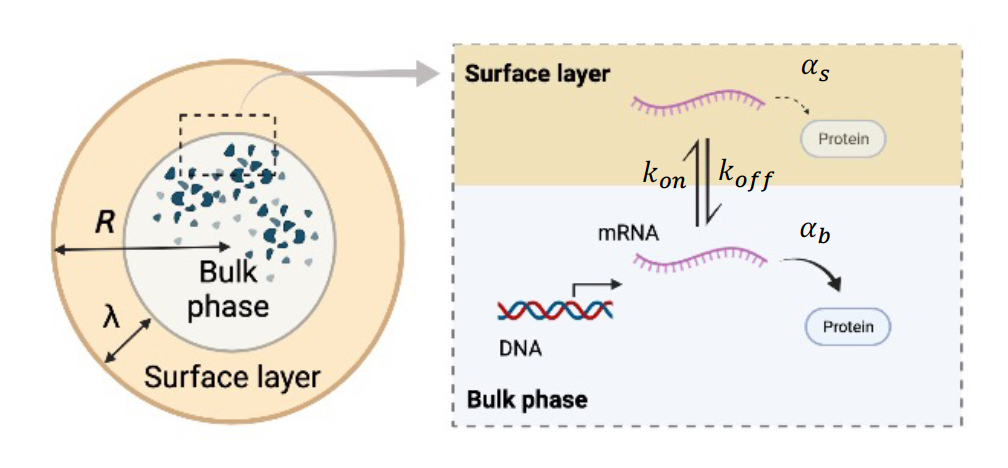}
\caption{\textbf{Schematic illustration of the TXTL reaction.} Gene expression is modeled considering the excluded volume effect in a cell-sized spherical compartment. In the bulk phase (yellow), the transcription at the reaction rate $\alpha_r$ and the translation at the reaction rate $\alpha_b$ proceed. On the other hand, in the surface layer (orange) transcription is completely suppressed. Furthermore, mRNA binds to boundary surface from the bulk phase and, in turn, the translation rate drops to $\alpha_s$ in the surface layer. Figure was created with BioRender.}\label{fig1}
\end{figure}

This section presents a mathematical model of a TXTL reaction encapsulated in a cell-sized space. For simplicity, we assume a spherical compartment of radius $R$, in which the molecular system for the TXTL reaction is enclosed (Fig. \ref{fig1}). We define $S$ and $V$ as the surface area and the volume of the confined spherical space, respectively ($S= 4\pi R^2$ and $V=\frac{4\pi}{3}R^3$). Among the molecules involved in gene expression, we assume that large protein complexes such as RNA polymerase and ribosomes (typical radius $R_g$) are subject to steric repulsion against the surface of the compartment. 

To consider the excluded volume effect of ribosomes near the boundary, we formulated the transcription reaction from DNA to mRNA and the translation reaction from mRNA to polypeptide protein in the two regions (Fig. \ref{fig1}). First, the surface layer is present beneath the compartment boundary with a thickness of $\lambda$ comparable to the radius $R_g$ of the large protein complexes involved in TXTL reactions. Typically, $\lambda$ is on the order of few tens nm, which is sufficiently small compared to radius $R$ ($\lambda \ll R$). Second, these protein complexes capture mRNA inefficiently inside the surface layer; thus, the translation of genetic information from mRNA into protein is likely to be suppressed. In contrast, the active TXTL reaction occurs in a bulk phase free from the excluded volume effect. For examining the size dependence of confined gene expression, we assumed that the concentration of transcriptional machinery (RNA polymerase) and translational machinery (ribosome) are the same in the compartments of different sizes.

Next, we consider the amount of the mRNA in the bulk region $N_b(t)$ and the amount of the mRNA in the surface layer $N_s(t)$ at time $t$. The rate at which mRNA in the bulk region attaches to the surface layer is defined as $k_{on} S\lambda$ with the binding rate per unit volume $k_{on}$, whereas the rate at which mRNA in the surface layer dissociates into the bulk region is defined as $k_{off} (V- S\lambda)$ with the detaching rate per unit volume $k_{off}$. The mRNA degradation rate is equal in both the bulk and surface layers, $\gamma_r$, and the transcription rate of mRNA is $\alpha_r V$. Based on the above reactions, the time evolution of $N_b(t)$ is
\begin{equation}\label{mRNA_number_bulk}
\frac{dN_b}{dt} = \alpha_r V- k_{on} S \lambda N_b(t)  + k_{off}(V - S\lambda) N_s(t) - \gamma_r N_b(t).
\end{equation}
For the analysis of geometric effects, the rescaled parameters and the description based on the mRNA concentration are useful. To this end, we rewrite Eq. \eqref{mRNA_number_bulk} with the mRNA concentration in the bulk region $r_b(t) = \frac{N_b(t)}{V}$, the mRNA concentration in the surface layer $r_s(t) = \frac{N_s(t)}{V}$, and the rescaled reaction rates  $k_{on} S \lambda \rightarrow k_{on}\frac{S\lambda}{V}$, $k_{off} (V-S\lambda) \approx k_{off} V \rightarrow k_{off}$ while the rate of mRNA synthesis $\alpha_r$ and the degradation rate $\gamma_r$ are scale-independent. Based on the above reactions, the time evolution of mRNA concentration in the bulk phase is rewritten as
\begin{equation}\label{mRNA_bulk}
\frac{dr_b}{dt} = \alpha_r - k_{on} \frac{S\lambda}{V} r_b(t) + k_{off} r_s(t) - \gamma_r r_b(t),
\end{equation}
where the surface layer to volume ratio of $\frac{S\lambda}{V} = \frac{3\lambda}{R}$ determines the capacity for mRNA in the surface layer in Eq \eqref{mRNA_bulk}. By taking the same formulation, the time evolution of mRNA in a surface layer is also given by
\begin{equation}\label{mRNA_surface}
\frac{dr_s}{dt} = k_{on} \frac{S\lambda}{V} r_b(t) - k_{off} r_s(t) - \gamma_r r_s(t)
\end{equation}

At the translation level, the mRNAs in each layer then serve as templates for the ribosomal translation process at a different protein production rate, $\alpha_b$ for the bulk phase, and $\alpha_s$ for the surface layer. The surface layer has a lower translation efficiency, that is, $\alpha_b \gg \alpha_s$. The average concentration of the protein synthesized in the compartment, $p(t)$, increases with time according to the following equation:
\begin{equation}\label{protein_ave}
\frac{dp}{dt} = \alpha_b r_b(t) + \alpha_s r_s(t) - \gamma_p p(t)
\end{equation}
where $\gamma_p$ is the degradation rate of expressed protein.

The focus of the present study was to reveal the size-dependence of the TXTL reactions at the steady state under confinement, and we then analyzed the steady-state concentrations of mRNA in each region ($\frac{dr_b}{dt} =0$, $\frac{dr_s}{dt} =0$) and that of the expressed protein ($\frac{dp}{dt} =0$). By solving Eq. \eqref{mRNA_bulk} and Eq. \eqref{mRNA_surface} with setting $\frac{dr_b}{dt}=0$ and $\frac{dr_s}{dt}=0$, the steady state concentrations of mRNA in bulk phase $\bar{r}_b$ and in the surface layer $\bar{r}_s$ are
\begin{equation}\label{mRNA_bulk_st}
\bar{r}_b = \frac{r_0}{1 + \frac{3\lambda}{R\tau}},
\end{equation}
and
\begin{equation}\label{mRNA_surf_st}
\bar{r}_s = \frac{3 \lambda}{R\tau} \frac{r_0}{1 + \frac{3\lambda}{R\tau}},
\end{equation}
where the mRNA concentration averaged over the compartment is $r_0 = \alpha_r/\gamma_r$, and the parameter of mRNA dissociation from the surface layer is $\tau = (k_{off} + \gamma_r)/k_{on}$. We can apply the same calculation to Eq. \eqref{protein_ave} to obtain the steady-state concentrations of protein summed over two regions $\bar{p}$ as follows:
\begin{equation}\label{protein_st}
\bar{p} = \frac{r_0}{\gamma_p}\frac{\alpha_b + \alpha_s \frac{3\lambda}{R\tau}}{1 + \frac{3\lambda}{R\tau}} .
\end{equation}
Eq. \eqref{protein_st} indicates the number of the expressed protein $N = \bar{p} V$ as
\begin{equation}\label{N_st}
N = \bar{p} V = \frac{r_0}{\gamma_p}\frac{4\pi R^3}{3}\frac{\alpha_b + \alpha_s \frac{3\lambda}{R\tau}}{1 + \frac{3\lambda}{R\tau}} .
\end{equation}

The dependence of $N$ on the constraint size $R$ is worth noting. The thickness of the surface layer $\lambda$ is a few tens of nanometers long, and if the size of the confinement is on the micron scale of a cell, $\lambda/R$ can be considered a minute amount. If the translation rate in the surface layer is suppressed ($\alpha_s \approx 0$), and the mRNA tends to dissociate from the surface layer ($\tau \gg 3\lambda/R$), then Eq. \eqref{N_st} is rewritten as
\begin{equation}\label{N_st1}
N \approx 
\frac{\alpha_b r_0}{\gamma_p}\frac{4\pi R^3}{3} \propto R^{3},
\end{equation}
meaning that the volume $V$ and the number of protein molecules $N$ follow the same size scaling, $V \propto R^3$, and $N \propto R^3$. Thus, the excluded volume effect in the surface layer during the translation process is almost negligible.

On the other hand, if the mRNA tends to be trapped in the surface layer ($\tau \ll 3\lambda/R$) and its translation is also significantly suppressed in the surface layer, the number of protein molecules is rewritten as
\begin{equation}\label{N_st2}
N \approx 
\frac{\alpha_b r_0}{\gamma_p}\frac{4\pi \tau R^4}{9\lambda} \propto R^4\lambda^{-1}.
\end{equation}
Eq. \eqref{N_st2} has a dependence on the constraint size $R$ of $N \propto R^4$, which is different from the scaling law of Eq. \eqref{N_st1} described above. This is because a large number of mRNAs are trapped by a factor of $R/\lambda$ in the translation-suppressing surface layer.

\begin{figure}[tbp]
\centering
\includegraphics[scale=0.26,bb=360 0 680 452]{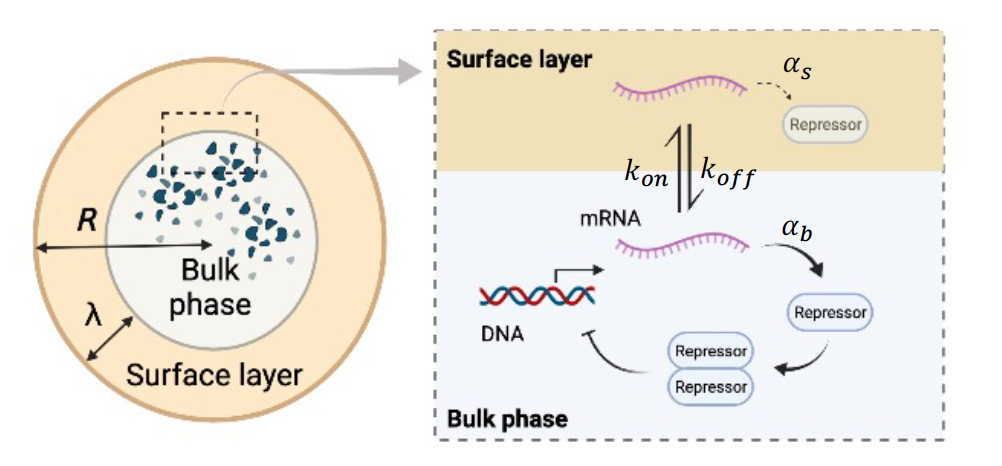}
\caption{\textbf{Schematic illustration of the TXTL reaction with negative autoregulatory feedback (NAF) control}. The repressor protein synthesized by the TXTL reaction forms a dimer (equilibrium constant $K_1$). This dimer binds to the operator region of DNA (equilibrium constant $K_2$). NAF control is realized by reducing the transcription rate of mRNA from the DNA in which the repressor dimer is bound. Figure was created with BioRender.}\label{fig2}
\end{figure}

\subsection*{Effect of negative autoregulatory feedback control}
The deviation from the ordinary size scaling shown in the previous section indicates that the TXTL reaction under confinement is affected by surface exclusion volume effects. In this section, we study whether such anomalous size scaling that originates from the encapsulation in cell-sized compartments is affected by transcriptional negative feedback control in confined TXTL reactions (Fig. \ref{fig2}). 

NAF control suppresses the production rate of mRNA at the transcriptional level. The expressed transcription repressor forms a dimeric complex, and the repressor dimer binds to the operator sequence. The complex of repressor dimer-operator DNA inhibits the process of mRNA synthesis, which achieves the repression of mRNA synthesis and, in turn, reduces the expression level of the transcriptional repressor protein (Fig. \ref{fig2}). The time evolution of the mRNA concentration in the bulk region under NAF control is described by the following equation: 
\begin{equation}\label{mRNA_bulk_neg}
\frac{dr_b}{dt} = \frac{\alpha_r}{1 + K_1K_2p(t)^2} - k_{on} \frac{S\lambda}{V} r_b(t) + k_{off} r_s(t) - \gamma_r r_b(t),
\end{equation}
where $K_1$ is the equilibrium constant for the dimerization of the transcription repressor, and $K_2$ is the equilibrium constant for the binding of the repressor dimer to the operator sequence in DNA. As for the transcription in the surface layer, mRNA can be present in the surface layer, but both transcription and translation beneath the boundary hardly occur due to the volume exclusion effect. Hence, the time evolution of the mRNA concentration in the surface layer follows the same equation as Eq. \eqref{mRNA_surface}.

The steady-state concentration of mRNA in the bulk region is 
\begin{equation}\label{mRNA_bulk_st_neg}
\bar{r}_b = \frac{r_0}{(1+ K_1K_2\bar{p}^2)(1 + \frac{3\lambda}{R\tau})},
\end{equation}
and the steady-state concentration of mRNA in the surface layer is
\begin{equation}\label{mRNA_surf_st_neg}
\bar{r}_s = \frac{3 \lambda}{R\tau} \frac{r_0}{(1+ K_1K_2\bar{p}^2)(1 + \frac{3\lambda}{R\tau})}.
\end{equation}

\begin{figure}[tb]
\centering
\includegraphics[scale=0.42,bb=0 0 550 452]{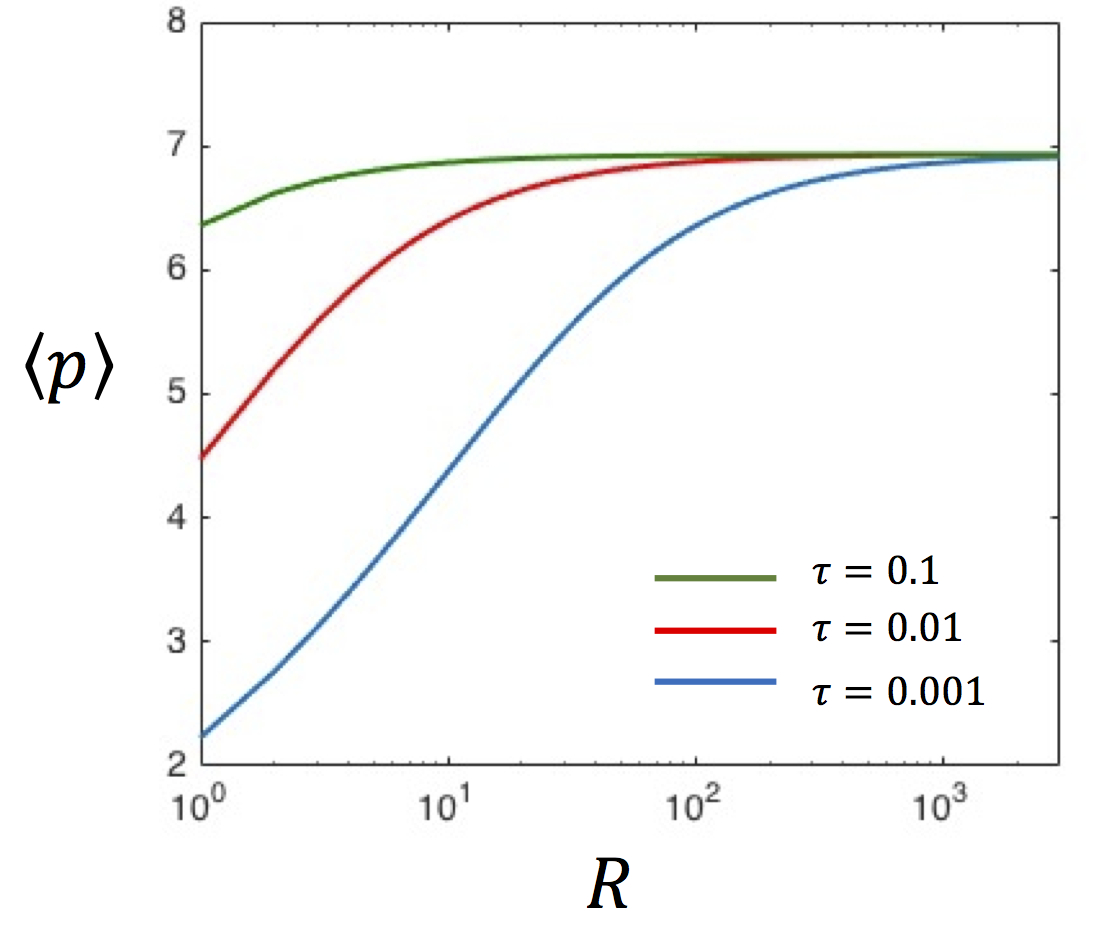}
\caption{\textbf{Steady state protein concentration $\bar{p}$ in the TXTL reaction with NAF control under the steric effect}.  $\bar{p}$ is plotted according to Eq. \eqref{protein_st_neg2} as a function of radius $R$. The three curves represent size dependence of $\bar{p}$ for each different parameter $\tau$ (Blue: $\tau = 0.001$, Red: $\tau = 0.01$, $\tau = 0.1$). The other parameters are $r_0=10$, $\gamma_p=0.03$, $\alpha_b=1$, $\alpha_s=0.001$, and $\lambda = 0.01$.}\label{fig3}
\end{figure}

Furthermore, for protein expression inside the compartment, the transcriptional repressor is degraded at the same rate $\gamma_p$ for the bulk region and the surface layer, and the translation from mRNA to protein follows the same equation as Eq. \eqref{protein_ave}. When gene expression is close to the steady state, the relation $K_ 1 K _2\bar{p}^2 \gg 1$ is established. Therefore, the reaction rate of the transcription can be approximated as $\frac{\alpha_r}{K_1 K 2\bar{p}^2}$. By solving for $\gamma_p\bar{p} = \alpha_b \bar{r}_b + \alpha_s \bar{r}_s$ using Eq. \eqref{mRNA_bulk_st_neg} and Eq. \eqref{mRNA_surf_st_neg} at the steady state yields
 
\begin{equation}\label{protein_st_neg2}
\bar{p} \approx \biggl(\frac{r_0}{K_1K_2\gamma_p}\frac{\alpha_b + \alpha_s \frac{3\lambda}{R\tau}}{1 + \frac{3\lambda}{R\tau}}\biggr)^{\frac{1}{3}} .
\end{equation}
Fig. \ref{fig3} shows the plot of steady-state protein concentration $\bar{p}$ against size $R$ based on Eq. \eqref{protein_st_neg2}. $\bar{p}$ drops at a small $R$. Such size-dependent reduction of protein concentration is similar to the TXTL reaction without NAF control (Eq. \eqref{protein_st}), but we need to further analyze Eq. \eqref {protein_st_neg2} to reveal the role of NAF control for the size-dependent TXTL reactions due to the excluded volume effect. 

As considered in Eq. \eqref{N_st1} in the previous section, when the translation rate in the surface layer is suppressed ($\alpha_s \approx 0$) and most of the mRNA is present in the bulk region ($\tau \gg 3\lambda/R$), the protein concentration at the steady state is $\bar{p} \approx \bigl(\frac{r_0}{K_1K_2\gamma_p} \alpha_b\bigr)^{\frac{1}{3}}$ based on Eq. \eqref{protein_st_neg2}. The number of protein molecules $N = \bar{p} V$ in the confined space is
\begin{equation}\label{N_st_neg1}
N \approx \frac{4\pi}{3} \biggl(\frac{r_0}{K_1K_2\gamma_p}\alpha_b\biggr)^{\frac{1}{3}}R^{3} \propto R^3.
\end{equation}
Similar to the case without NAF control, we find that the volume $V$ and the number of protein molecules $N$ follow the same size scaling $V \propto R^3$ and $N \propto R^{3}$. Regular size scaling relation is maintained because the excluded volume effect in the surface layer is almost negligible. 
In contrast, when mRNA tends to stay in the surface layer ($\tau \ll 3\lambda/R$) and NAF control undergoes under the influence of the strong excluded volume effect, the protein concentration can be evaluated as $\bar{p} \approx \Bigl[\frac{r_0}{K_1K_2\gamma_p}\bigl(\alpha_s + \alpha_b\frac{R\tau}{3\lambda}\bigr)\Bigr]^{\frac{1}{3}}$. The number of expressed proteins is 
\begin{equation}\label{N_st_neg2}
N \approx  \frac{4\pi}{3}\biggl[\frac{r_0}{K_1K_2\gamma_p}\Bigl(\alpha_b\frac{R\tau}{3\lambda}\Bigr)\biggr]^{\frac{1}{3}}R^3 \propto R^{\frac{10}{3}}\lambda^{-\frac{1}{3}}.
\end{equation}
The newly obtained size scaling in Eq. \eqref{N_st_neg2} differs from the scaling law of Eq. \eqref{N_st2} at the case without NAF control. This analysis implies that NAF control alleviates the anomalous volume scaling originating from the excluded volume effect and transforms it into a closely normal size scaling of $R^{10/3}$ on the change in compartment size. Without NAF control, a doubling of the volume, such as in cell division, would double the molecular concentration due to gene expression. However, with NAF control, the change in protein concentration at the two-fold cell volume was limited to $2^{1/3} \approx 1.26$ times. This analysis suggests that the NAF control can limit the possible change in molecular concentration arose from the excluded volume effect to a small variation.

\subsection*{Effect of positive autoregulatory feedback control}

\begin{figure}[tbp]
\centering
\includegraphics[scale=0.4,bb=30 0 680 292]{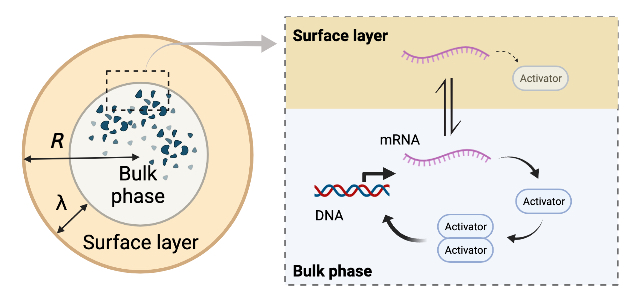}
\caption{\textbf{Schematic illustration of the TXTL reaction with positive autoregulatory feedback (PAF) control}. The activator protein synthesized by the TXTL reaction forms a dimer at equilibrium constant $K_3$. The activator dimer binds to the operator region at equilibrium constant $K_4$. PAF control upregulates mRNA synthesis. Figure was created with BioRender.}\label{fig4}
\end{figure}

Towards the construction of artificial bioreactors, another important mechanism for transcriptional regulation is positive autoregulatory feedback control, which is another network motif widely identified in transcriptional circuits \cite{serrano2, collins2, maeda3}. Gene networks with PAF control confers multistability in cell-fate decision. Using the similar model of confined gene expression as in Eqns \eqref{mRNA_bulk}-\eqref{protein_st_neg2}, we next ask a question what size dependence would be observed if the gene expression reaction with PAF control inside a small compartment (Fig. \ref{fig4}). 

PAF control activates the production rate of mRNA at the transcriptional level. The expressed transcription activator forms a dimeric complex that can bind to the operator sequence. The activator dimer-operator DNA recruits RNA polymerase close to the promoter region. RNA polymerase proceeds the mRNA synthesis and, in turn, increases the expression level of the activator protein at the translational level (Fig. \ref{fig4}). For simplicity, we assume that transcription does not occur from a promoter in which the activator protein is not bound to the operator region.

The time evolution of the mRNA concentration in the bulk region under PAF control is described by the following equation: 
\begin{equation}\label{mRNA_bulk_pos}
\frac{dr_b}{dt} =\alpha_r \frac{K_3K_4p(t)^2}{1 + K_3K_4p(t)^2} - k_{on} \frac{S\lambda}{V} r_b(t) + k_{off} r_s(t) - \gamma_r r_b(t),
\end{equation}
where $K_3$ is the equilibrium constant for the dimerization of the transcription activator, and $K_4$ is the equilibrium constant for the binding of the activator dimer to the operator sequence. 

The steady-state concentration of mRNA in the bulk region is 
\begin{equation}\label{mRNA_bulk_st_pos}
\bar{r}_b = r_0\frac{K_3K_4\bar{p}^2}{(1+ K_3K_4\bar{p}^2)(1 + \frac{3\lambda}{R\tau})},
\end{equation}
and the steady-state concentration of mRNA in the surface layer is
\begin{equation}\label{mRNA_surf_st_pos}
\bar{r}_s = r_0\frac{3 \lambda}{R\tau} \frac{K_3K_4\bar{p}^2}{(1+ K_3K_4\bar{p}^2)(1 + \frac{3\lambda}{R\tau})}.
\end{equation}

Near the steady state, where gene expression occurs and the amount of activator protein is maximally increased, the relationship $ K_3K_4\bar{p}^2 \gg 1$ holds. Then, the steady-state concentration of mRNA is approximated by the same equation as in the absence of feedback. Thus, in a gene expression response regulated by PAF, as the compartment size decreases, the size-dependence is $N \propto R^4\lambda^{-1}$ that is the same anomalous scaling as seen in the gene circuit without autoregulatory feedback Eq. \eqref{N_st2}. This result also has implications for the fact that gene expression by NAFs is an effective mode of regulation that exhibits size-dependent repression.

\subsection*{Cooperative negative autoregulatory feedback control}
Moreover, autoregulatory feedback becomes highly nonlinear by changing the cooperative multimer formation of transcription factors. In the above calculations, the multimer formation was limited to the dimer complex, but our model can be extended to an association reaction where one transcriptional complex is made from n monomers. Consider that the multimerized transcriptional complex binds to the operator sequence and represses its transcriptional activity. In this case, the time evolution of the mRNA concentration in the bulk region under cooperative NAF control is
\begin{equation}\label{mRNA_bulk_neg_nth}
\frac{dr_b}{dt} = \frac{\alpha_r}{1 + K_{1n}K_{2n}p(t)^n} - k_{on} \frac{S\lambda}{V} r_b(t) + k_{off} r_s(t) - \gamma_r r_b(t),
\end{equation}
where $K_{1n}$ is the equilibrium constant for the multimerization of the transcription repressor, and $K_{2n}$ is the equilibrium constant for the binding of the multimer complex to the operator sequence.

The steady-state concentration of mRNA in the bulk region is also calculated by taking the same approach,
\begin{equation}\label{mRNA_bulk_st_neg_nth}
\bar{r}_b = \frac{r_0}{(1+ K_{1n}K_{2n}\bar{p}^n)(1 + \frac{3\lambda}{R\tau})},
\end{equation}
and the steady-state concentration of mRNA in the surface layer is
\begin{equation}\label{mRNA_surf_st_neg_nth}
\bar{r}_s = \frac{3 \lambda}{R\tau} \frac{r_0}{(1+ K_{1n}K_{2n}\bar{p}^n)(1 + \frac{3\lambda}{R\tau})}.
\end{equation}

By solving $\gamma_p\bar{p} = \alpha_b \bar{r}_b + \alpha_s \bar{r}_s$ at the steady state, the steady state concentration of the protein product is
\begin{equation}\label{protein_st_neg2_nth}
\bar{p} \approx \biggl(\frac{r_0}{K_{1n}K_{2n}\gamma_p}\frac{\alpha_b + \alpha_s \frac{3\lambda}{R\tau}}{1 + \frac{3\lambda}{R\tau}}\biggr)^{\frac{1}{n+1}}.
\end{equation}
When the reaction is suppressed in the surface layer at a small compartment size ($\alpha_s \approx 0$, $\tau \ll 3\lambda/R$), the number of expressed proteins $\bar{p} V$ is 
\begin{equation}\label{N_st_neg2_nth}
N \approx  \frac{4\pi}{3}\biggl[\frac{r_0}{K_{1n}K_{2n}\gamma_p}\Bigl(\alpha_b\frac{R\tau}{3\lambda}\Bigr)\biggr]^{\frac{1}{n+1}}R^3 \propto R^{3+\frac{1}{n+1}}\lambda^{-\frac{1}{n+1}}.
\end{equation} 
As cooperativity $n$ increases, the amount of protein inside the compartment is close to the normal size scaling $R^3$. This means that having NAF control and making its feedback highly nonlinear is a promising strategy to reduce the excluded volume effect regardless of the compartment size.

\section*{Discussion}

\begin{figure}[tbp]
\centering
\includegraphics[scale=0.13,bb=1510 0 280 692]{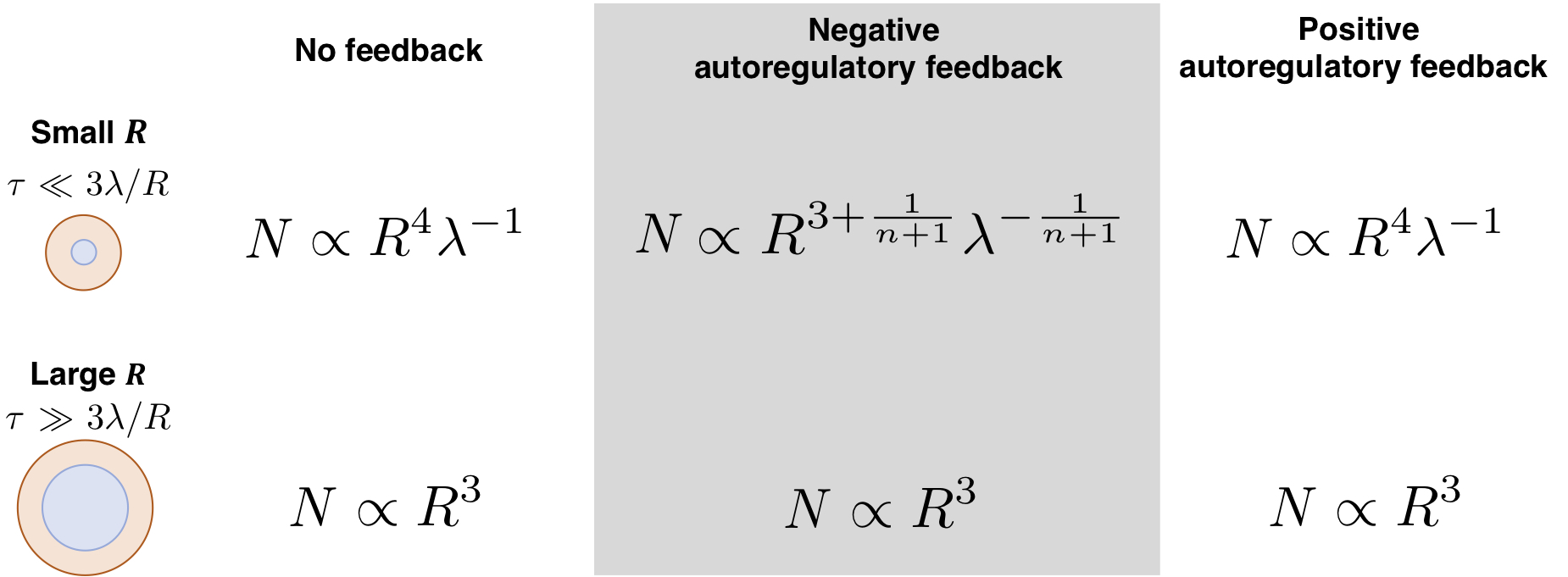}
\caption{\textbf{Confinement dependence of the TXTL reaction with autoregulatory feedback controls}. The size dependence of the amount of expressed protein $N$ in the compartment is shown for small and large radial size $R$. The cooperativity of multimer formation is $n$-th order.}\label{fig5}
\end{figure}

In recent years, cell-free extracts have been used to study biological phenomena \cite{noireaux1,jewett,noireaux2,noireaux3,macdonald,hibi}. The exclusion volume effect near the membrane boundary becomes more prominent as the cell size decreases; thus, the confinement effect cannot be ignored. A previous study found that in simple gene expression, the exclusion volume effect near the interface contributes to repression at the translation level, which changes the size-dependent scaling of the compartment at small droplet sizes \cite{sakamoto}. The present study extends this to a system with transcriptional autoregulatory feedback, showing that the NAF control of transcriptional regulation is an effective scaling law that avoids the anomalous size-dependent scaling law. The mathematical model in this study has demonstrated the size-dependent scaling relation of gene expression in a microcellular environment and that the scaling exponent can be changed by NAF control (Fig. \ref{fig5}). The transcriptional NAF control suppresses the amount of protein synthesis in the bulk region and prevents excessive production. Because the suppressive excluded volume effect originating from the surface layer coexists with the repressive gene regulation originating from feedback control in bulk, indicating that the TXTL reaction is suppressive throughout the entire area in the compartment, making it difficult for anomalous size dependence to appear. In fact, in tiny bacteria such as \textit{E. coli}, network motifs in which negative feedback regulation frequently appear in transcriptional circuits are known \cite{alon1,alon2}. Thus, the NAF control is an essential regulatory mechanism that contributes to the suppression of size-dependent fluctuations among heterogeneous cell populations. 

Our theoretical model shows the effects of steric confinement on gene expression, including transcriptional feedback, mainly in small compartments, such as those in bacteria and artificial bioreactors. The bulk phase corresponds to the cytoplasm in bacteria, and the surface layer is a thin region near the cell membrane, where the large ribosome complex is excluded because of its size. In small prokaryotes and artificial cell systems smaller than \SI{10}{\micro\meter}, the excluded volume effect on the surface is not negligible \cite{sakamoto}. In contrast, the typical size of eukaryotic cells is approximately \SI{100}{\micro\meter} \cite{milo}, which is larger than the compartment size assumed in this study. In addition, as eukaryotic cells have ribosomes located on the endoplasmic reticulum membrane, where protein translation occurs, the excluded volume effect of macromolecules is expected to have only a relatively small effect. Such differences may also be the crucial difference between prokaryotes and eukaryotes in their attempts to suppress size-dependence in gene expression.

Finally, experimental verification of our theoretical predictions is an important challenge for the future. Placing the cI gene (encoding lambda repressor CI) downstream of the P$_{R}$ promoter would result in a circuit with negative autoregulatory feedback \cite{maeda3}. This theoretical prediction can be examined by measuring the amount of CI protein with a GFP probe in W/O droplets of various sizes. Furthermore, similar genetic circuit would change to have positive autoregulatory feedback if a cI lambda repressor is placed downstream of the P$_{RM}$ promoter \cite{collins2,maeda3}. Therefore, the scaling relationship of compartment volume and the amount of protein product would be able to indicate whether negative autoregulatory feedback is the mechanism that mitigates the suppression of gene expression in a confined space.

\subsection*{Acknowledgements}
The author thanks K. Kaya for conducting the preliminary experiment that was the basis of this study. Graphic illustration was created with BioRender software. This work was supported by Grant-in-Aid for Scientific Research on Innovative Areas ``Molecular Engines" JP18H05427, Grant-in-Aid for Scientific Research (B) JP20H01872, Grant-in-Aid for Challenging Research JP21K18605.

\subsection*{Data availability}
All data generated or analyzed during this study are included in this published article.

\subsection*{Author contributions}
Y.T.M. designed the research, constructed mathematical model, analyzed the model, and wrote the manuscript.

\subsection*{Competing interests}
The author declares no competing interests.

\end{document}